\documentclass{article}
\usepackage{spconf,amsmath,graphicx,balance}
\usepackage{amssymb,amsfonts}
\usepackage{algorithmic}
\usepackage{graphicx,tikz,pgfplots}
\usepackage{textcomp}
\usepackage[colorlinks=true,  pdfstartview=FitV,
linkcolor=green, citecolor=green, urlcolor=green]{hyperref}

\def\F{{\mathcal{F}}}
\def\E{{\mathcal{E}}}

\title{Improving Acquisition Speed  of X-Ray Ptychography through \\ Spatial Undersampling and Regularization}
\makeatletter

\def\@name{\em{Prasan Shedligeri$^{1*}$, Florian Schiffers$^3$, Semih Barutcu$^{4}$,  Pablo Ruiz$^2$,} \\ \emph{Aggelos K. Katsaggelos$^4$, Oliver Cossairt$^{3,4}$} 
\thanks{This work was supported in part by the US Department of Energy through the Los Alamos National Laboratory. Los Alamos National Laboratory is operated by Triad National Security, LLC, for the National Nuclear Security Administration of U.S. Department of Energy (Contract No. 89233218CNA000001). The work was also funded in part by NSF CAREER IIS-1453192. The research is based in part upon work supported by the Office of the Director of National Intelligence (ODNI), Intelligence Advanced Research Projects Activity (IARPA), via IARPA-16003-165043 contract number. The views and conclusions contained herein are those of the authors and should not be interpreted as necessarily representing the official policies or endorsements, either expressed or implied, of the ODNI, IARPA, or the U.S. Government. The U.S. Government is authorized to reproduce and distribute reprints for Governmental purposes notwithstanding any copyright annotation thereon. Prasan Shedligeri was supported by a Research Travel Scholarship from Robert Bosch Center for Data Science and AI, IIT Madras, India.}
}
\address{$^1$Department of Electrical Engineering, IIT Madras, India \hspace{5pt} $^2$OriGen.AI, Brooklyn, NY 11201, USA\\
$^3$Department of Computer Science, Northwestern University, Evanston, IL 60208, USA \\
$^4$Department of Electrical and Computer Engineering, Northwestern University, Evanston, IL 60208, USA\\
\\
{\small $^*$ ee16d409@ee.iitm.ac.in}} 
\begin{document}
%
\maketitle
\begin{abstract}
X-ray ptychography is one of the versatile techniques for nanometer resolution imaging. 
The magnitude of the diffraction patterns is recorded on a detector and the phase of the diffraction patterns \textcolor{black}{is} estimated using phase retrieval techniques.
Most phase retrieval algorithms make the solution well-posed by relying on the constraints imposed by the overlapping region between neighboring diffraction pattern samples. 
As the overlap between neighboring diffraction patterns reduces, the problem becomes ill-posed and the object cannot be recovered. 
To avoid the ill-posedness, we investigate the effect of regularizing the phase retrieval algorithm with image priors for various overlap ratios between the neighboring diffraction patterns. 
We show that the object can be faithfully reconstructed at low overlap ratios by regularizing the phase retrieval algorithm with image priors such as Total-Variation and Structure Tensor Prior.
We also show the effectiveness of our proposed algorithm on real \textcolor{black}{data acquired from an IC chip} with a coherent X-ray beam.

\end{abstract}

\section{Introduction}
\vspace{-10pt}
\label{sec:intro}
X-ray ptychography has become one of the most popular techniques to achieve resolutions of up to a few nanometers in imaging objects.
The ptychography technique was first proposed by Hoppe~\cite{hoppe1969beugung} in 1969 and was then further developed experimentally and algorithmically in the 2000s~\cite{thibault2009probe,maiden2009improved,guizar2008phase,faulkner2004movable}.
This technique has found applications in both research and industry in fields such as biological imaging and \textcolor{black}{material} sciences.
A focused, coherent X-ray probe beam interacts with the object to be imaged and produces an exit wavefront. 
The exit wavefront propagates to the far-field and the propagated wavefront is approximated by the Fourier transform of the exit wavefront.
A 2D detector is placed in the far-field which records the intensities of the propagated wavefront.
As the probe beam is narrow, it interacts with only a small part of the object each time, and the probe is then scanned across the entire object in steps.
Each time, the detector only records the square of the magnitude of the Fourier transform of the wavefront exiting from the object whereas the phase is completely lost.
Therefore, recovering the object field requires solving an ill-posed phase-retrieval problem, where one has to reconstruct the complex object field given only the magnitude of its Fourier transform~\cite{marchesini2007invited,elser2003phase,fienup1982phase,gerchberg1972practical}.
To \textcolor{black}{force} the phase retrieval problem to have a unique solution, sufficient overlap between successive scan points of the probe beam is necessary~\cite{bunk2008influence}.

Scanning the object with enough overlap between successive scan points typically leads to very long acquisition \textcolor{black}{times and inhibits} high throughput scanning.
In the field of image processing, ill-posed problems are typically regularized by imposing prior knowledge about the image to be restored. 
Several image priors have been proposed in the literature ranging from signal-agnostic priors to signal-specific priors.
In this work, we propose to regularize the solution to the phase retrieval algorithm by imposing constraints arising from the prior knowledge about the object to be estimated.
We explore the use of regularizers that are well suited for the reconstruction of integrated circuits (IC) from X-Ray diffraction measurements.

\begin{figure*}
    \centering
    \includegraphics[width=0.8\textwidth]{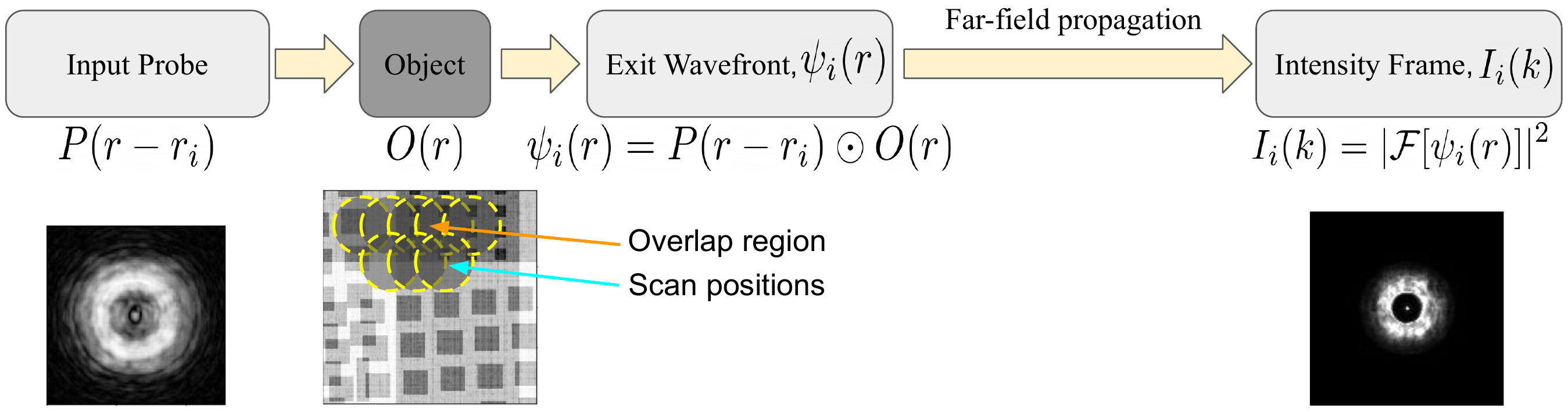}
    \vspace{-10pt}
    \caption{Ptychographic object acquisition model. A complex probe beam $P$ interacts with a complex object $O$ to produce an exit wavefront $\psi$. The wavefront $\psi$ propagates to the far field where the magnitude of the wavefront is recorded on the sensor.}
    \label{fig:intro}
    \vspace{-15pt}
\end{figure*}

Many phase retrieval algorithms attempt to iteratively estimate the object by imposing the constraints observed in the overlap region and the measurement observed in the diffraction patterns. 
Many gradient descent based algorithms have also been proposed which rely on manually derived closed-form expressions for the gradient of the cost-function~\cite{ marchesini2007invited,zhong2016nonlinear,candes2015phase}.
Very recently, alternative algorithms have been introduced which perform gradient descent using automatic differentiation of the cost-function~\cite{ghosh2018adp,kandel2019using,nashed2017distributed}.
These algorithms have the flexibility of using complex, differentiable terms in the cost function and can also provide easy hardware acceleration on the GPU.
We adopt the automatic differentiation based ptychographic phase retrieval algorithm~\cite{ghosh2018adp,kandel2019using} for investigating the role of image prior in regularizing the ill-posedness.
Specifically, we investigate two different image prior models a) Total-Variation (TV)~\cite{rudin1992nonlinear} and b) Structure Tensor Prior~(STP)~\cite{lefkimmiatis2013convex}.
We show that the 2D object fields of the IC circuits can be faithfully reconstructed at low overlap ratios by regularizing the phase retrieval algorithm with image priors such as Total-Variation~\cite{rudin1992nonlinear} and Structure Tensor Prior~(STP)~\cite{lefkimmiatis2013convex}. 
It has been previously shown that data-driven prior models can help regularize the ill-posed phase retrieval problem when the overlap ratio is low~\cite{boominathan2018phase,shamshad2019deep} in Fourier ptychography.
Here, we investigate the impact of using various prior models for regularizing the phase retrieval problem for X-ray ptychography under various overlap ratios. 
We note that, in addition to reconstructing the object field, our ptychography algorithm also estimates the complex probe beam as well.
\vspace{-10pt}
\section{Regularized X-ray Ptychography}
\vspace{-10pt}
In X-ray ptychography, a focused, coherent X-ray probe beam $P(r)$ interacts with a complex object $O(r)$.
The \textcolor{black}{X-ray beam exits} the object as a wavefront $\psi(r)$ and then propagates to the far-field which can be approximated by the Fourier-transform.
A detector is placed at the far-field which records the diffraction pattern $I(k)$ of the far-field propagated wavefront $\psi(r)$.
Mathematically, the ptychographic imaging process is modeled as,
\begin{align}
    \vspace{-8pt}
    \psi_i(r) &= P(r - r_i) \odot O(r)
    \label{eq:forward}\\
    I_i(k) &= |\F[\psi_i(r)]|^2
    \label{eq:magnitude}
    \vspace{-8pt}
\end{align}
where $r_i$ is the current position of the coherent probe beam, $\odot$ indicates pointwise multiplication, and $\mathcal{F}$ denotes the Fourier-transform from the real-space ($r$) to the reciprocal-space ($k$).
\textcolor{black}{In this problem, the quantities of interest to be estimated are the probe beam $P(r)$ and the complex object $O(r)$.
Let $\hat P(r)$ and $\hat O(r)$ be the estimates of the original signal.
Following the work of Ghosh~\emph{et al.}~\cite{ghosh2018adp}, we formulate the estimation of these quantities as the minimization of the objective function defined as }
\begin{equation}
    \vspace{-8pt}
    \E_o = \frac{1}{M}\sum_{i\in M} \{|\mathcal{F}[\hat P(r - r_i) \odot \hat O(r)]| - \sqrt{I_i(k)}\}^2
    \label{eq:objective}
    \vspace{-5pt}
\end{equation}
where $M$ is the number of probe positions. 
In each iteration, a batch of $M$ probe \textcolor{black}{positions is sampled} and the objective $\E_o$ is computed according to eq.~\eqref{eq:objective}. 
The energy $\E_o$ is minimized using gradient descent steps where the gradients are obtained using an automatic differentiation algorithm~\cite{ghosh2018adp,kandel2019using}.

\vspace{-10pt}
\subsection{Object Regularization}
\vspace{-8pt}
In this work, we mainly consider ptychographic imaging of IC chips with nanometer resolution.
We know for a fact that such objects when imaged \textcolor{black}{have a piecewise} smoothness property and hence we consider prior models that impose such constraint on the estimated phase and magnitude of the object.
Particularly, we consider two prior models a) Total Variation (TV) \cite{rudin1992nonlinear} and b) Structure Tensor Prior~(STP)~\cite{lefkimmiatis2013convex}. 
As the object $O$ to be estimated is a complex field, we impose the prior models separately on the object-magnitude $|O|$ and the object-phase $\angle O$.
\textcolor{black}{TV priors have} been widely used in the field of image processing and is defined as 
\begin{equation}
\mathcal{E}_p^{TV} = \sum_{i} \|\nabla_i |O|\|_2 + \sum_{i} \|\nabla_i \angle O\|_2
\label{eq:TV_prior}
\vspace{-5pt}
\end{equation}
where $\nabla_i$ is the finite difference image gradient operator.
STP\cite{lefkimmiatis2013convex} imposes a sparsity constraint on the eigenvalues of the structure tensor $S_K\textbf{I}$.
The structure tensor $S_K\textbf{I}$ is defined as $S_K\textbf{I} = K * \{H(\textbf{I})\}$, 
where $K$ is a gaussian smoothing kernel, $*$ denotes convolution and $H(\textbf{I})$ is the pixel-wise Hessian of the image $\textbf{I}$. 
Let $\lambda_i^+ \geq \lambda_i^- \geq 0$ be the eigen values of $S_K(I)$, then the cost for the STP is defined as,
\begin{equation}
    \mathcal{E}_p^{STP} = \frac{1}{N} \sum_{i=1}^N |\lambda_i^+| + |\lambda_i^-|
    \label{eq:STP_prior}
    \vspace{-5pt}
\end{equation} 

Besides regularizing the phase retrieval algorithm with image prior we also propose to exploit correlations between the phase and the magnitude estimations.
To do that, we employ the cross-channel prior proposed by Heide~\emph{et al.}~\cite{heide2013high}.
The cross-channel prior~\cite{heide2013high} over the object magnitude $|O|$ and the object-phase $\angle O$ is defined as
\begin{align}
    \E_p^{CC} = \|~\nabla \angle O ~\odot~ |O|~ - ~\nabla |O| ~\odot~ \angle O~ \|_1
    \label{eq:cross_channel}
\end{align}
where $\nabla$ denotes the finite difference gradient operator and $\odot$ represents element-wise multiplication. 

\vspace{-10pt}
\subsection{Probe Retrieval}
In a ptychographic imaging experiment, it is difficult to know the exact complex beam profile of the probe $P(r)$.
In \cite{ghosh2018adp}, the authors initialize the probe $P^0$ as
\begin{align}
    P^0 = \F^{-1} \Big\{\frac{1}{M} \sum_{i=1}^{M} \sqrt{I_i(k)} ~\Big\},
    \label{eq:probeInit}
\end{align}
and update it after each iteration of the object reconstruction algorithm.
We follow a similar approach to estimate the probe initialization as in eq.~\eqref{eq:probeInit}.
Additionally, the probe $P^0$ is Fresnel propagated by a few millimeters to obtain a defocused \textcolor{black}{probe.}
We also know that the magnitude of the beam profile varies smoothly across the probe.
We define a cost $\E_p^{pr}$ that enforces a smoothness constraint on the probe magnitude to our overall objective. 
We define $\E_p^{pr} = \|~ \nabla|P|~ \|_2$, 
where $|P|$ denotes the magnitude of the estimated probe.
Overall our objective becomes,
\begin{align}
    O,P = \arg\min_{O,P}~~ \E_o + \lambda_1 \E_p^x + \lambda_2 \E_p^{CC} + \lambda_3 \E_p^{pr}
    \vspace{-5pt}
\end{align}
where $x\in\{TV, STP\}$ and $\lambda_1$, $\lambda_2$ and $\lambda_3$ are the regularization hyper-parameters.

\vspace{-10pt}
\section{Experiments}
\vspace{-10pt}
In our experiments we consider a complex object, a projection of a chip, as shown in Fig.~\ref{fig:probe_sm}a) for simulating the diffraction patterns. 
The probe $P$ is set to be $256\times256$ pixels, whose magnitude is a gaussian kernel with a standard deviation of $64$ pixels.
The phase of the probe is obtained from a reconstruction of a real object using the algorithm proposed in \cite{nashed2017distributed}.

We initialize the optimization algorithm with each pixel of object magnitude and phase uniformly drawn between $[0.9,1.0]$ and \textcolor{black}{$[-0.1, 0.1]$,} respectively.
The probe is initialized 
as described in eq.~\eqref{eq:probeInit}. 
We use Adam optimizer~\cite{kingma2014adam} with a batch size of $16$, an initial learning rate of $0.1$ for the object, and $0.01$ for the probe.
The optimization algorithm is implemented with the Pytorch~\cite{paszke2019pytorch} automatic differentiation framework.

\vspace{-10pt}
\subsection{Probe smoothness}
\begin{figure}[t!]
  \centering
  \setlength{\tabcolsep}{0.1em} 
  \begin{tabular}{ccccccc}
    
         & \footnotesize{Ground Truth} & \footnotesize{Cropped} & \footnotesize{No Prior}  & \footnotesize{Cropped} & \footnotesize{$\E_p^{pr}$}  & \footnotesize{Cropped} \\
         
            \rotatebox{90}{\hspace{22pt}\scriptsize Phase} &
        \includegraphics[trim={0 0.8cm 1.3cm 0.6cm},clip,width=0.2\columnwidth]{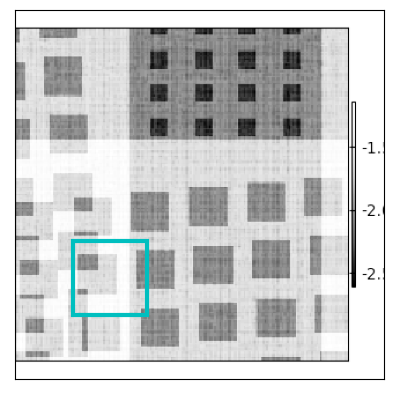} &
        \includegraphics[width=0.1\columnwidth]{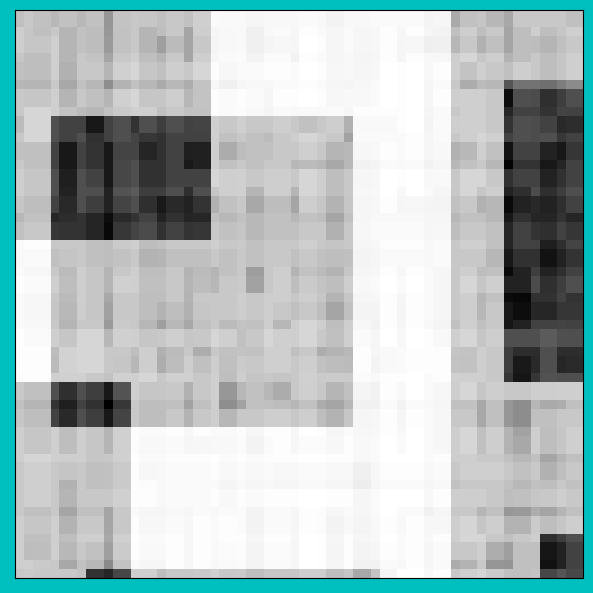} &
        \includegraphics[trim={0 0.8cm 1.3cm 0.6cm},clip,width=0.2\columnwidth]{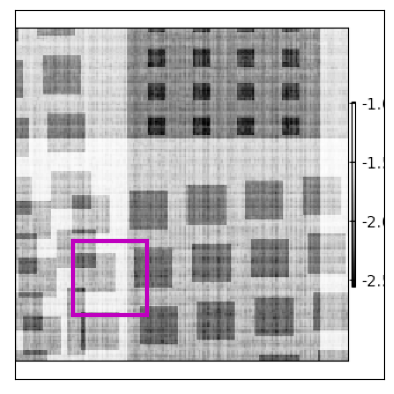} & 
        \includegraphics[width=0.1\columnwidth]{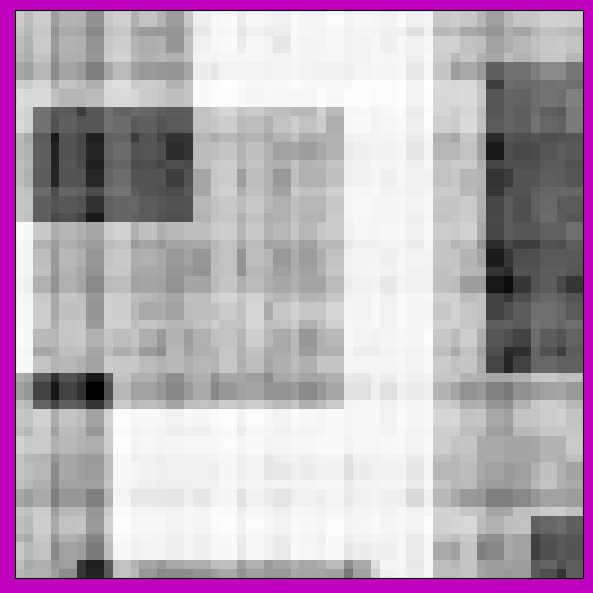} &
        \includegraphics[trim={0 0.8cm 1.3cm 0.6cm},clip,width=0.2\columnwidth]{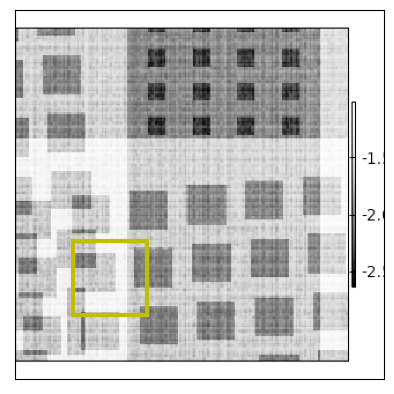} &
        \includegraphics[width=0.1\columnwidth]{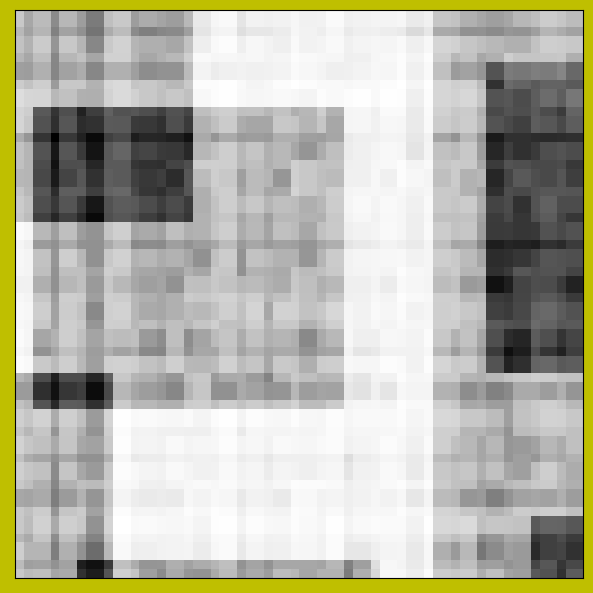}\\

    \rotatebox{90}{\hspace{15pt}\scriptsize Magnitude} &
        \includegraphics[trim={0 0.8cm 1.3cm 0.6cm},clip,width=0.2\columnwidth]{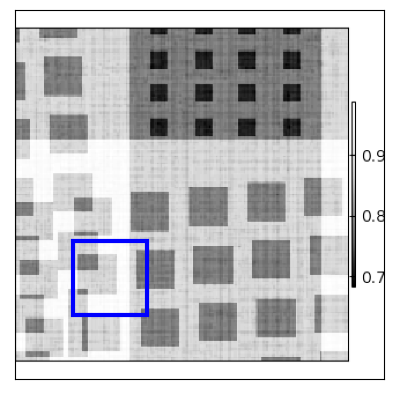} &
        \includegraphics[width=0.1\columnwidth]{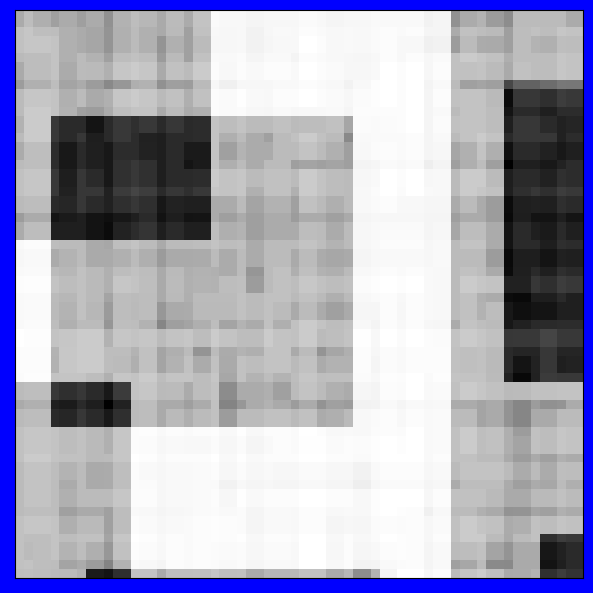} &
        \includegraphics[trim={0 0.8cm 1.3cm 0.6cm},clip,width=0.2\columnwidth]{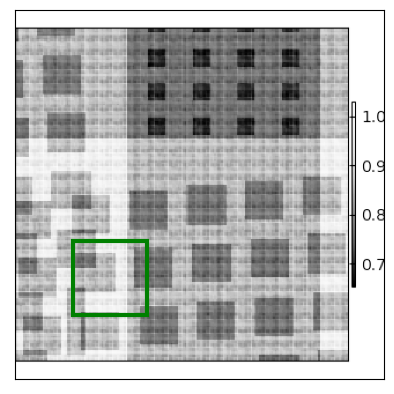} &
        \includegraphics[width=0.1\columnwidth]{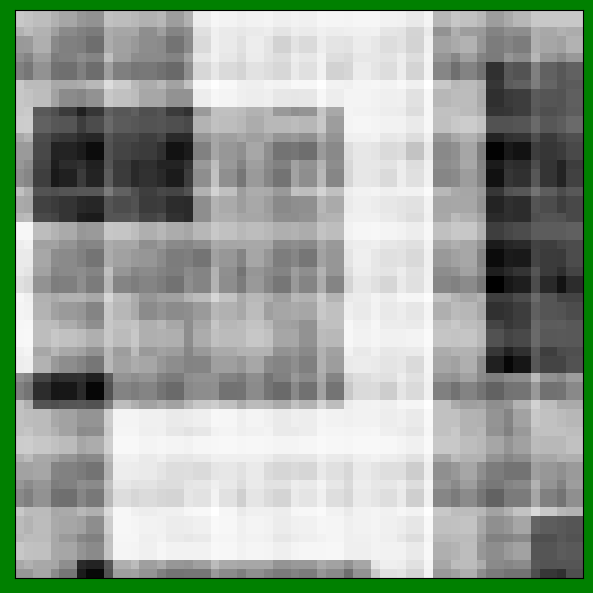} & 
        \includegraphics[trim={0 0.8cm 1.3cm 0.6cm},clip,width=0.2\columnwidth]{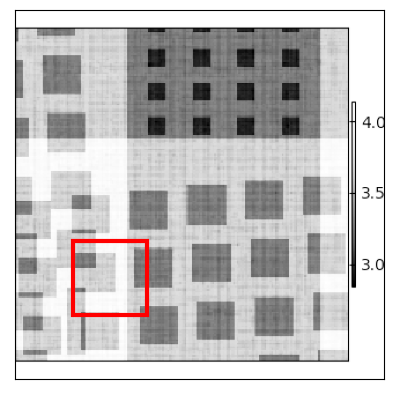} & 
        \includegraphics[width=0.1\columnwidth]{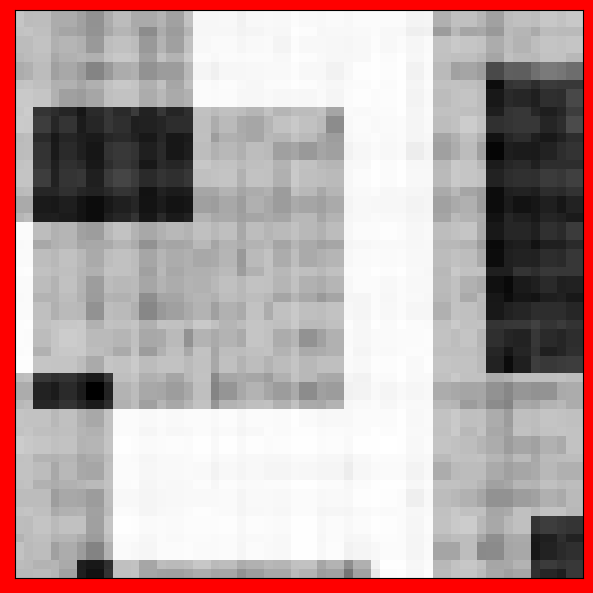}\\
              & (a) &  & (b) & & (c) & 
   \end{tabular}
    \caption{We compare the magnitude and phase of the complex object field estimated without and with probe smoothness loss at an overlap ratio of $78\%$.}
   \label{fig:probe_sm}
   \vspace{-10pt}
   \end{figure}

\vspace{-5pt}
\label{sub:probe}
Here, we investigate the effectiveness of imposing the probe smoothness prior on the estimated probe.
We simulate the diffraction patterns with a step size of $32$ pixels which translates to an overlap of $78\%$.
For the case of probe smoothness loss, we minimize the objective $\E = \E_o + \lambda \E_p^{pr}$ with $\lambda=0.01$.
In Fig.~\ref{fig:probe_sm}b) and \ref{fig:probe_sm}c), we show the reconstructed object without and with the probe \textcolor{black}{smoothness loss, respectively}.
We observe the object estimated when the probe smoothness loss is more smooth and both the object magnitude and phase are \textcolor{black}{free from periodic} artifacts.

    \begin{figure}[t]
\centering
\setlength{\tabcolsep}{0.3em} 

  \begin{tabular}{ccccc}
    
         & \footnotesize{Ground Truth} & \footnotesize{No Prior}   & \footnotesize{$\E_p^{pr}$}  & \footnotesize{$\E_p^{pr}+\E_p^{CC}$}  \\
         
            \rotatebox{90}{\hspace{22pt}\scriptsize Phase} &
        \includegraphics[trim={0 0.5cm 1.0cm 0.5cm},clip,width=0.2\columnwidth]{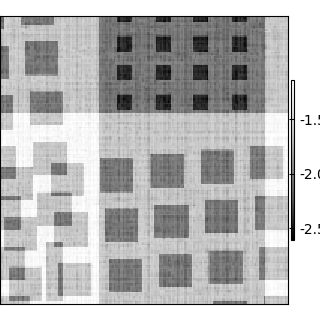} &
        \includegraphics[trim={0 0.5cm 1.0cm 0.5cm},clip,width=0.2\columnwidth]{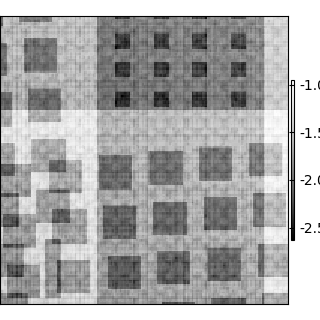} &   
        \includegraphics[trim={0 0.5cm 1.0cm 0.5cm},clip,width=0.2\columnwidth]{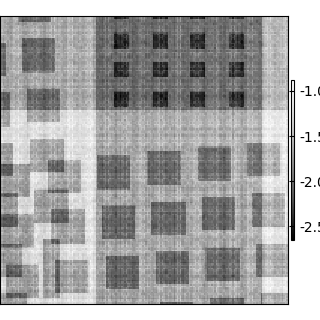} & 
        \includegraphics[trim={0 0.5cm 1.0cm 0.5cm},clip,width=0.2\columnwidth]{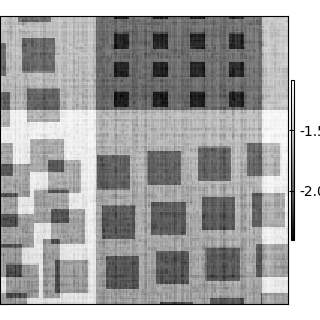}\\
    
    \rotatebox{90}{\hspace{15pt}\scriptsize Magnitude} &
      \includegraphics[trim={0 0.5cm 1.0cm 0.5cm},clip,width=0.2\columnwidth]{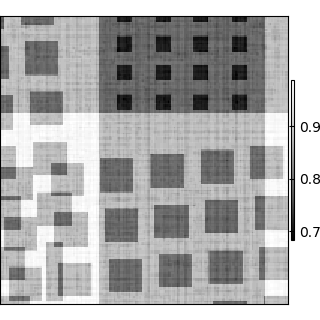} &   
      \includegraphics[trim={0 0.5cm 1.0cm 0.5cm},clip,width=0.2\columnwidth]{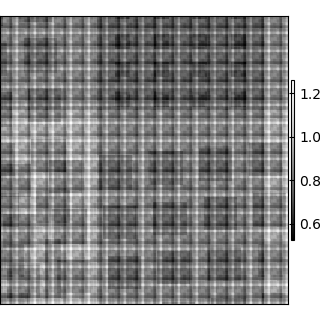} &   
      \includegraphics[trim={0 0.5cm 1.0cm 0.5cm},clip,width=0.2\columnwidth]{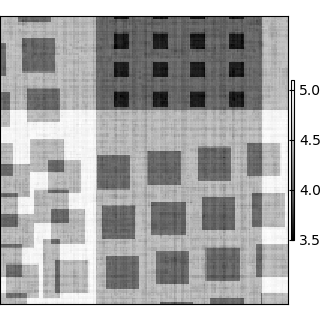} & 
      \includegraphics[trim={0 0.5cm 1.0cm 0.5cm},clip,width=0.2\columnwidth]{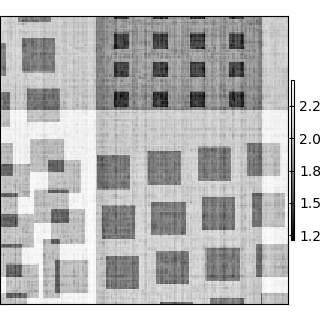}\\
     
   \end{tabular}
   \vspace{-10pt}
    \caption{In the above figure, we compare the complex object field estimated with an overlap ratio of $57\%$.}
   \label{fig:Cross-channel}
   \vspace{-10pt}
\end{figure}

\begin{figure}[t!]
   \centering
    \setlength{\tabcolsep}{0.3em}
  \begin{tabular}{ccccc}

    & \footnotesize{78\% overlap} & \footnotesize{57\% overlap} & \footnotesize{36\% overlap} & \footnotesize{15\% overlap} \\

    \rotatebox{90}{\hspace{16pt}\scriptsize $\E_p^{CC}$} &
    \includegraphics[trim={0 0.5cm 1.0cm 0.5cm},clip,width=0.2\columnwidth]{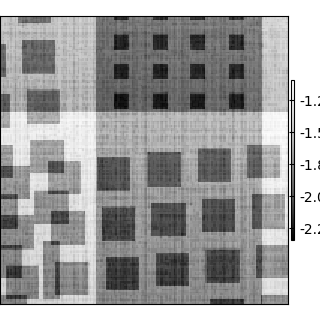} &   
    \includegraphics[trim={0 0.5cm 1.0cm 0.5cm},clip,width=0.2\columnwidth]{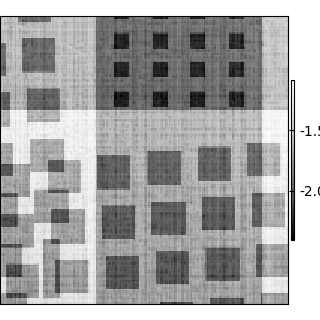} &   
    \includegraphics[trim={0 0.5cm 1.0cm 0.5cm},clip,width=0.2\columnwidth]{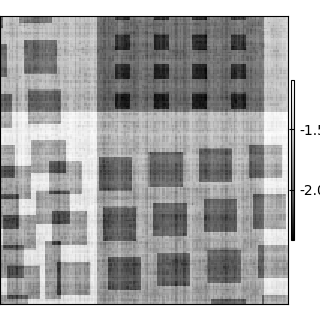} &
    \includegraphics[trim={0 0.5cm 1.0cm 0.5cm},clip,width=0.2\columnwidth]{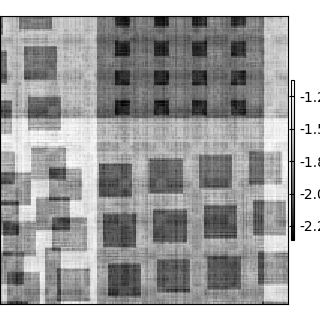} \\

    \rotatebox{90}{\hspace{05pt}\scriptsize $\E_p^{CC}+\E_p^{TV}$} &
    \includegraphics[trim={0 0.5cm 1.0cm 0.5cm},clip,width=0.2\columnwidth]{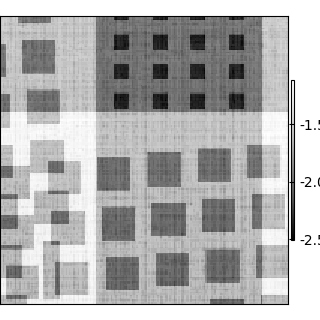} &   
    \includegraphics[trim={0 0.5cm 1.0cm 0.5cm},clip,width=0.2\columnwidth]{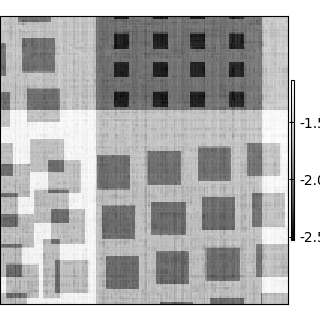} &   
    \includegraphics[trim={0 0.5cm 1.0cm 0.5cm},clip,width=0.2\columnwidth]{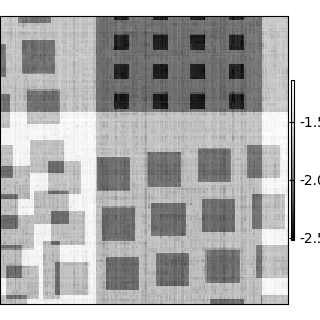} &
    \includegraphics[trim={0 0.5cm 1.0cm 0.5cm},clip,width=0.2\columnwidth]{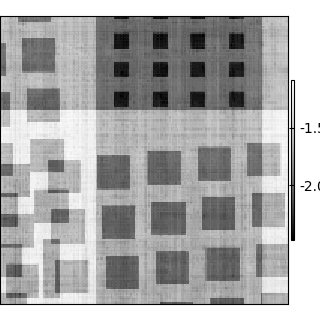} \\

    \rotatebox{90}{\hspace{02pt}\scriptsize $\E_p^{CC}+\E_p^{STP}$} &
    \includegraphics[trim={0 0.5cm 1.0cm 0.5cm},clip,width=0.2\columnwidth]{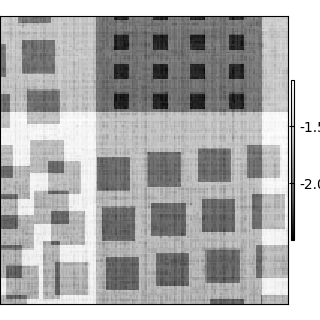} &   
    \includegraphics[trim={0 0.5cm 1.0cm 0.5cm},clip,width=0.2\columnwidth]{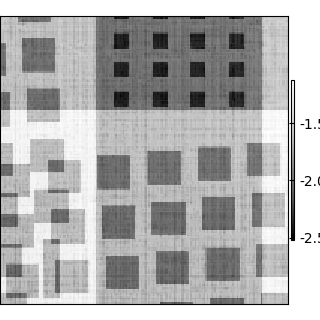} &   
    \includegraphics[trim={0 0.5cm 1.0cm 0.5cm},clip,width=0.2\columnwidth]{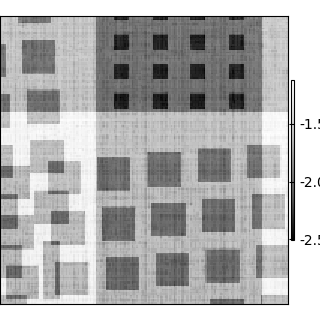} &
    \includegraphics[trim={0 0.5cm 1.0cm 0.5cm},clip,width=0.2\columnwidth]{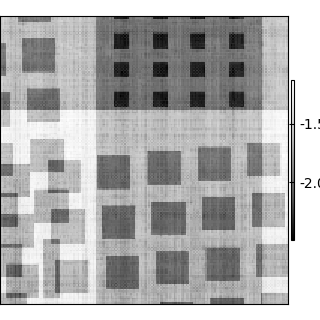} \\
\vspace{-10pt}
\end{tabular}
    \caption{We show the estimated object under various overlap ratios while imposing the different prior models.
    We observe that while the cross-channel prior does well for high overlap, it fails for very low overlap ratios.
    With gradient sparsity inducing priors such as TV and STP object can be recovered well for even very low overlap ratios.}
   \label{fig:TV_STP}
   \vspace{-15pt}
\end{figure}

    

\subsection{Cross channel (CC) prior}
\vspace{-5pt}
To investigate the effectiveness \textcolor{black}{of the cross-channel} prior we simulate diffraction patterns with a step size of $64$ pixels, leading \textcolor{black}{to an overlap} ratio of $57\%$. 
We compare the effectiveness \textcolor{black}{of the cross-channel} prior with that of using \textcolor{black}{only the probe-smoothness} prior and using no prior at all.
For the case of probe smoothness loss, we minimize the same object as in Sec.~\ref{sub:probe} and for the case of the cross-channel prior we minimize the objective $\E = \E_o + \lambda_1 \E_p^{pr} + \lambda_2 \E_p^{CC}$ where $\lambda_1 = \lambda_2 = 0.01$.
In Fig.~\ref{fig:Cross-channel}, we show the reconstructed object using no prior, probe smoothness prior alone and using both the probe smoothness prior as well as the cross-channel prior.
We observe that the cross-channel prior regularizes the solution well and estimates an artifact-free object.

\vspace{-10pt}
\subsection{Gradient Sparsity Regularization}
\begin{figure}[t]
   \centering

\begin{tikzpicture}[scale=0.5]
\begin{axis}[
legend cell align={left},
legend style={at={(0.03,0.03)}, anchor=south west, draw=white!80.0!black},
tick align=outside,
tick pos=left,
x grid style={white!69.01960784313725!black},
xlabel={Overlap Ratio},
xmin=10, xmax=80,
x dir=reverse,
xtick style={color=black},
y grid style={white!69.01960784313725!black},
ylabel={SSIM},
ymin=0.7, ymax=1,
ytick style={color=black}
]
\addplot [semithick, blue, mark=*, mark size=3, mark options={solid}]
table {%
78 0.89
57 0.86
36 0.82
15 0.79
};
\addlegendentry{No Prior}
\addplot [semithick, red, mark=triangle*, mark size=3, mark options={solid}]
table {%
78 0.89
57 0.86
36 0.91
15 0.88
};
\addlegendentry{TV Prior}
\addplot [semithick, green!50.0!black, mark=asterisk, mark size=3, mark options={solid}]
table {%
78 0.89
57 0.87
36 0.89
15 0.87
};
\addlegendentry{STP}

\end{axis}
\end{tikzpicture}
  \vspace{-15pt}
   \caption{Here, we show the plot of SSIM of the estimated object phase for different overlap ratios when the object is estimated without any prior, TV prior and the STP prior.}
   \label{fig:ssimVSoverlap}
   \setlength{\tabcolsep}{0.3em}
  \begin{tabular}{ccccc}
    & \footnotesize{EPIE~\cite{maiden2009improved}} & \footnotesize{No Prior} & \footnotesize{$\E_p^{CC}+\E_p^{TV}$} & \footnotesize{$\E_p^{CC}+\E_p^{STP}$}\\
            \rotatebox{90}{\hspace{4pt}\scriptsize $175$ samples} &
        \includegraphics[trim={0 0.8cm 1.3cm 0.6cm},clip,width=0.18\columnwidth]{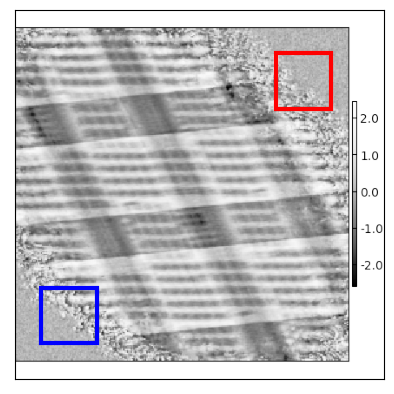} &   
        \includegraphics[trim={0 0.8cm 1.3cm 0.6cm},clip,width=0.18\columnwidth]{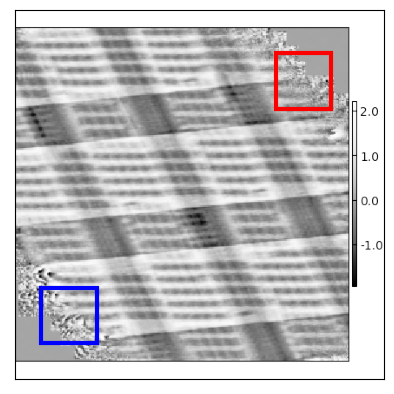} &   
        \includegraphics[trim={0 0.8cm 1.3cm 0.6cm},clip,width=0.18\columnwidth]{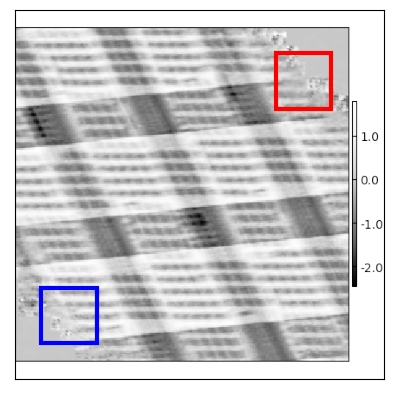}&
        \includegraphics[trim={0 0.8cm 1.3cm 0.6cm},clip,width=0.18\columnwidth]{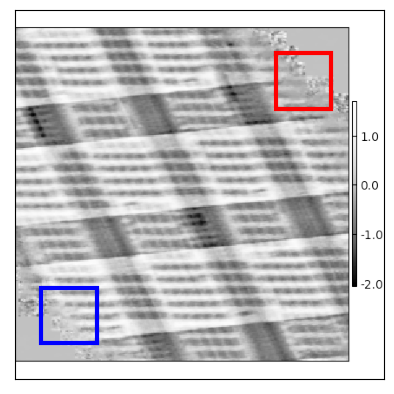} \\
        
         &
        \includegraphics[trim={0 0.4cm 0 0.4cm},clip,angle=0,width=0.18\columnwidth]{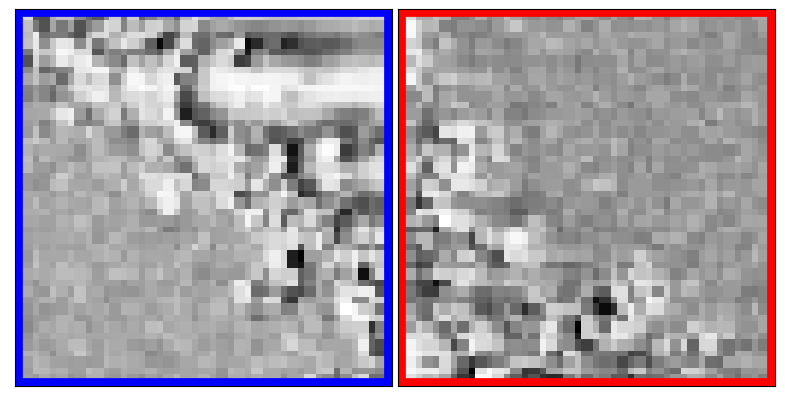} &   
        \includegraphics[trim={0 0.4cm 0 0.4cm},clip,angle=0,width=0.18\columnwidth]{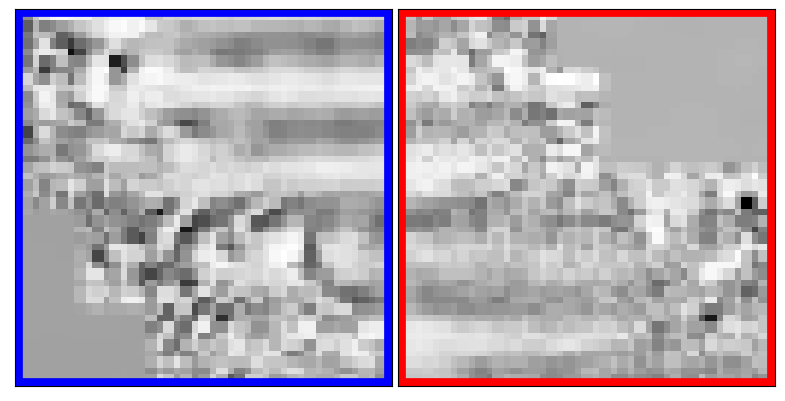} &   
        \includegraphics[trim={0 0.4cm 0 0.4cm},clip,angle=0,width=0.18\columnwidth]{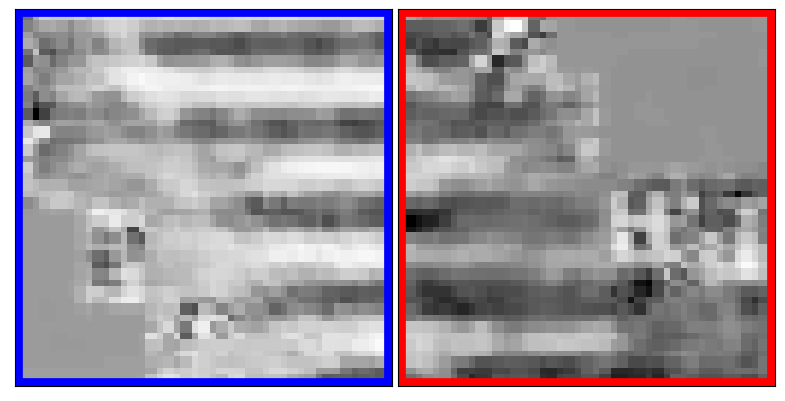}&
        \includegraphics[trim={0 0.4cm 0 0.4cm},clip,angle=0,width=0.18\columnwidth]{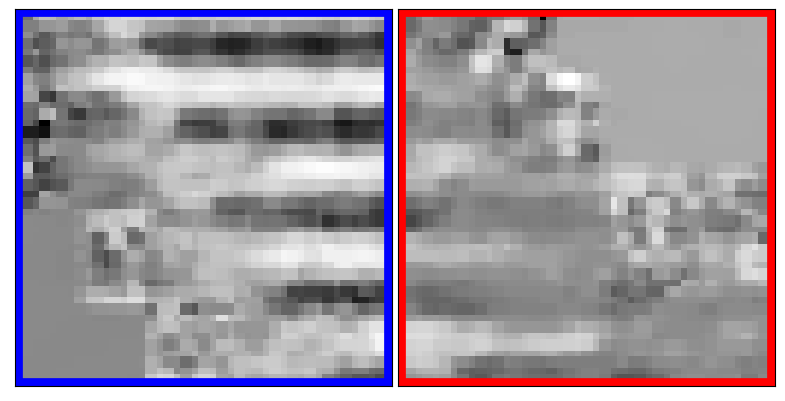} \\
        
            \rotatebox{90}{\hspace{4pt} \scriptsize $99$ samples} &
        \includegraphics[trim={0 0.8cm 1.3cm 0.6cm},clip,width=0.18\columnwidth]{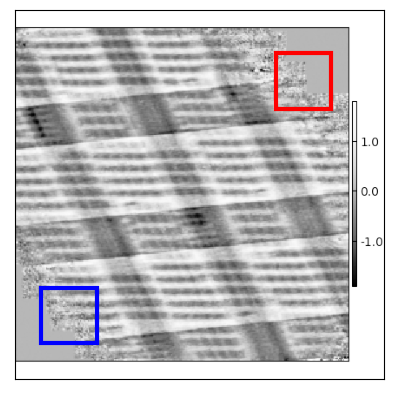} &   
        \includegraphics[trim={0 0.8cm 1.3cm 0.6cm},clip,width=0.18\columnwidth]{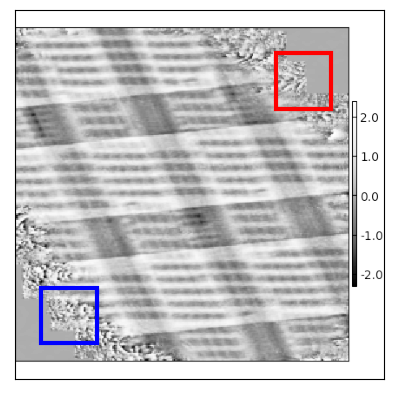} &   
        \includegraphics[trim={0 0.8cm 1.3cm 0.6cm},clip,width=0.18\columnwidth]{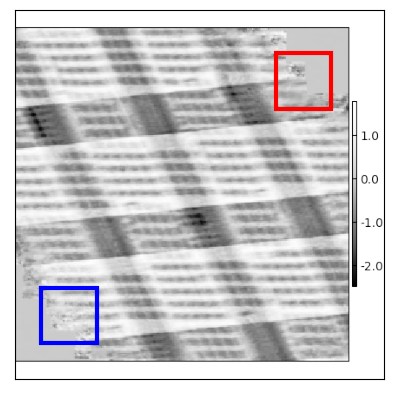}&
        \includegraphics[trim={0 0.8cm 1.3cm 0.6cm},clip,width=0.18\columnwidth]{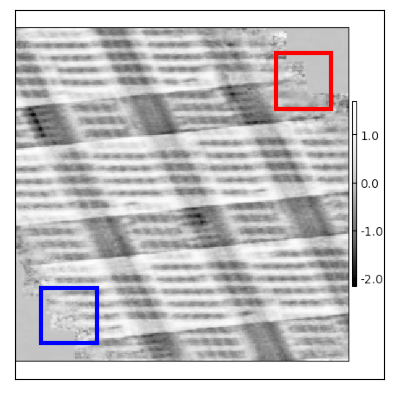} \\
        
         &
        \includegraphics[trim={0 0.4cm 0 0.4cm},clip,angle=0,width=0.18\columnwidth]{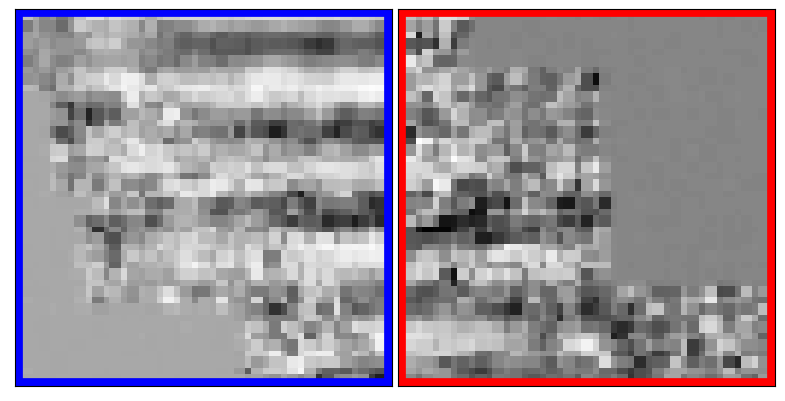} &   
        \includegraphics[trim={0 0.4cm 0 0.4cm},clip,angle=0,width=0.18\columnwidth]{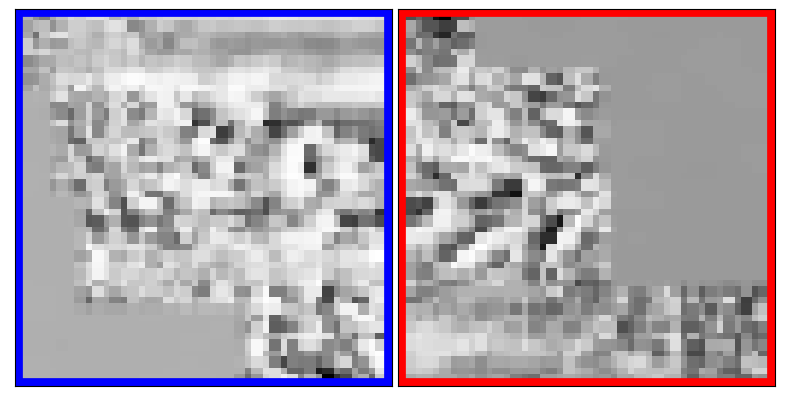} &   
        \includegraphics[trim={0 0.4cm 0 0.4cm},clip,angle=0,width=0.18\columnwidth]{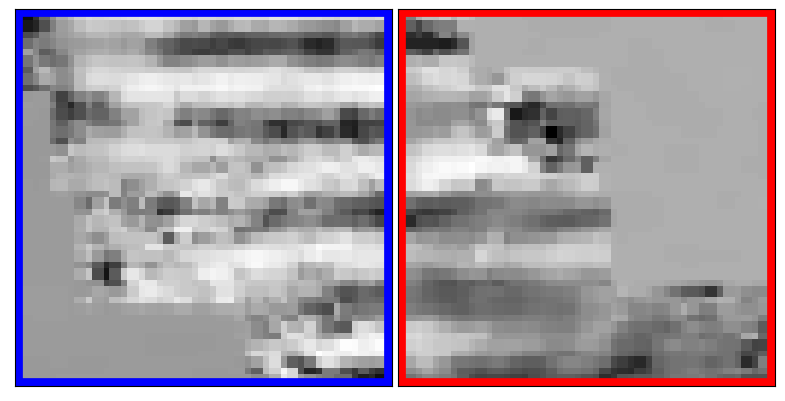}&
        \includegraphics[trim={0 0.4cm 0 0.4cm},clip,angle=0,width=0.18\columnwidth]{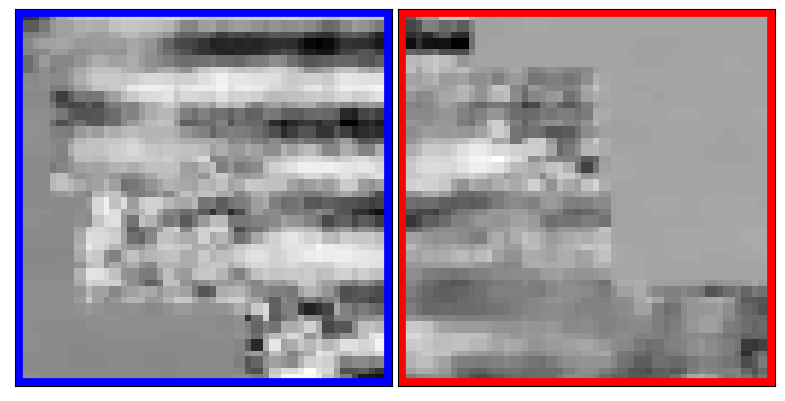} \\
        
            \rotatebox{90}{\hspace{4pt} \scriptsize  $61$ samples} &
        \includegraphics[trim={0 0.8cm 1.3cm 0.6cm},clip,width=0.18\columnwidth]{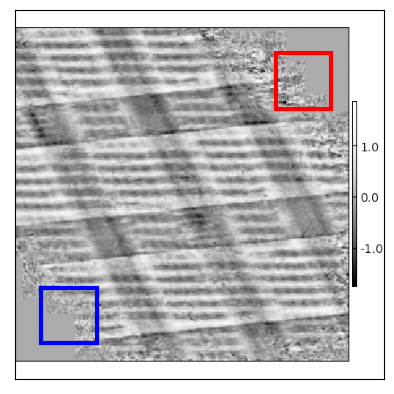} &
        \includegraphics[trim={0 0.8cm 1.3cm 0.6cm},clip,width=0.18\columnwidth]{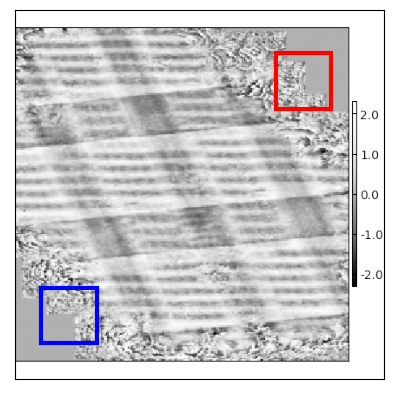} &   
        \includegraphics[trim={0 0.8cm 1.3cm 0.6cm},clip,width=0.18\columnwidth]{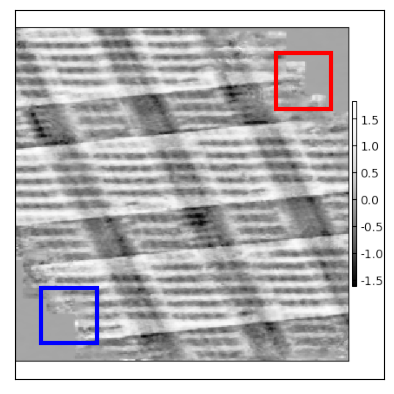}&
        \includegraphics[trim={0 0.8cm 1.3cm 0.6cm},clip,width=0.18\columnwidth]{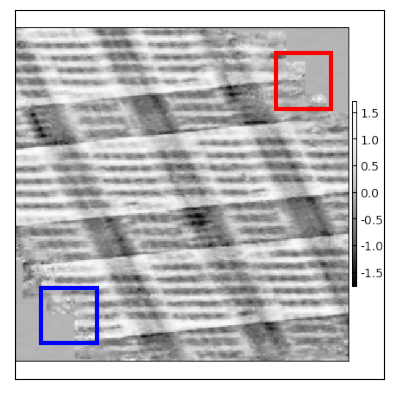} \\
        
        &
        \includegraphics[trim={0cm 0.4cm 0.1cm 0.4cm},clip,angle=0,width=0.18\columnwidth]{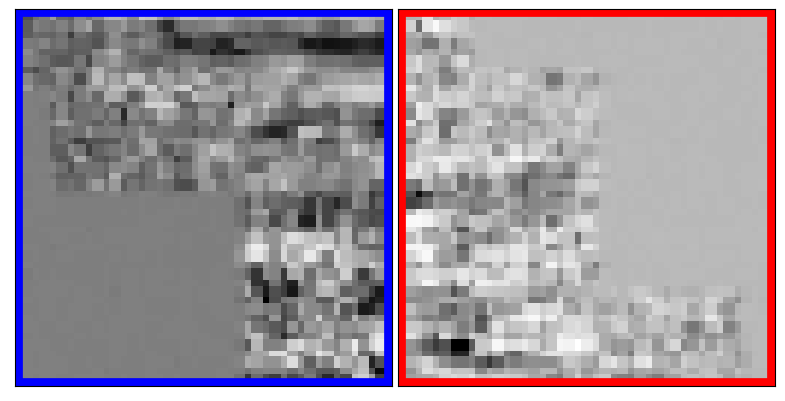} &   
        \includegraphics[trim={0cm 0.4cm 0.1cm 0.4cm},clip,angle=0,width=0.18\columnwidth]{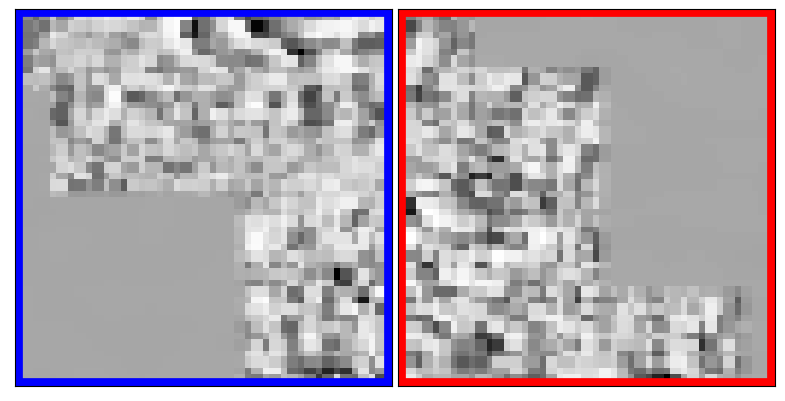} &   
        \includegraphics[trim={0cm 0.4cm 0.1cm 0.4cm},clip,angle=0,width=0.18\columnwidth]{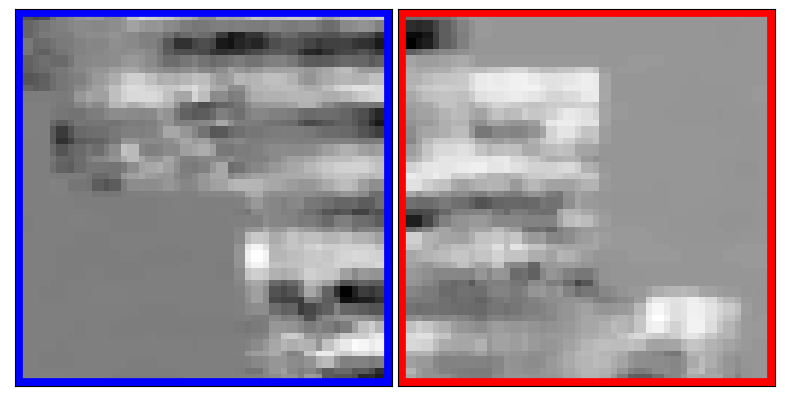}&
        \includegraphics[trim={0cm 0.4cm 0.1cm 0.4cm},clip,angle=0,width=0.18\columnwidth]{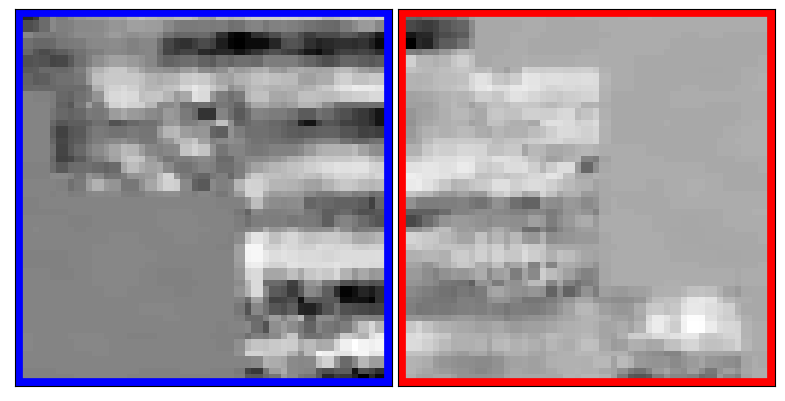} \\
        \vspace{-10pt}
   \end{tabular}
    \caption{We show the phase of the complex object reconstructed using the captured real data. With higher step sizes (indicating low overlap ratios) the object estimate degrades rapidly when no prior is imposed. When using the TV or the STP prior the object is recovered well for low overlap ratios as well.}
   \label{fig:realData}
   \vspace{-15pt}
\end{figure}

\vspace{-5pt}
Here, we investigate the effect of imposing gradient sparsity \textcolor{black}{on the} estimated object. 
Here, we compare two different prior models a) TV prior and b) Structure Tensor Prior.
We minimize the objective $\E = \E_o + \lambda_1 \E_p^{pr} + \lambda_2 \E_p^{CC} + \lambda_3 \E_p^x$ where $x=TV$ or $x=STP$. 
We simulate the diffraction patterns for $4$ different overlap ratios of $78\%$, $57\%$, $36\%$ and $15\%$. 
We set $\lambda_1=\lambda_2=0.01$ for object reconstruction for all the overlap ratios. 
We set $\lambda_3 = 0.005$ for overlap ratios of $78\%$, $57\%$ and $\lambda_3=0.01$ for overlap ratios of $36\%$ and $15\%$.
We show the reconstructed objects with TV prior and STP prior in Fig.~\ref{fig:TV_STP}.
For comparison, we also show the objects reconstructed \textcolor{black}{with the cross-channel} prior as well for all the overlap ratios.
We observe \textcolor{black}{that the cross-channel} prior fails to regularize the solution at very low overlap ratios.
However, gradient sparsity inducing priors such as TV and STP can recover the major details of the object at very low overlap ratios.
We also provide a quantitative comparison of the estimated objects in Fig~\ref{fig:ssimVSoverlap}.
At very low overlap ratios these priors over-smooth the reconstructed objects and cannot recover the fine details.

\vspace{-10pt}
\subsection{Real data}
\vspace{-5pt}
We demonstrate the effectiveness of the proposed regularization methods on real data that were obtained from ~\cite{holler2019three}. 
The data consists of $175$ diffraction patterns, recorded in the region of interest of the IC chip following the Fermat spiral trajectory~\cite{huang2014optimization}.
The scan points have an average spacing of $0.7\mu m$.
We artificially introduce undersampling in the scan obtained from the object by holding out some of the captured diffraction patterns from the reconstruction algorithm.
We consider three cases where the object is reconstructed using a) all the $175$ diffraction patterns, b) only $99$ diffraction patterns and c) only $61$ diffraction patterns.
The undersampling is done by uniformly thinning the set of $175$ 2-D scan points.
We show the object reconstructed from the above three cases without \textcolor{black}{using a prior, using a TV prior and using an STP prior} in Fig.~\ref{fig:realData}. 
\textcolor{black}{We also show the complex object reconstructed under various overlap ratios from a previously proposed phase retrieval algorithm EPIE~\cite{maiden2009improved} in Fig.~\ref{fig:realData}.
EPIE~\cite{maiden2009improved} is an iterative greedy algorithm which simultaneously reconstructs both the probe and the object and does not impose any signal prior constraint on the estimated quantities.
When the signal prior is not imposed during reconstruction, both our algorithm and EPIE quickly degrade in the reconstruction performance with decreasing overlap ratios.
}
When using the STP and the TV priors the algorithm is able to recover most of the details even for very low overlap ratios.

\vspace{-12pt}
\section{Conclusions}
\vspace{-8pt}
High throughput scanning in X-ray ptychography requires the overlap between the neighboring scan points to be low.
Here, we investigate the ill-posedness of the phase retrieval algorithm for object recovery for various overlap ratios.
We introduce regularization of the phase retrieval algorithm by imposing various prior models such as TV, STP, and CC priors.
We show that using prior models in the minimization objective can regularize the phase retrieval for very low overlap ratios.
When the overlap is high, prior models help in removing the artifacts compared to not using any prior model.
For very low overlap ratios, the prior models tend to over-smooth the reconstructed object and hence better and stronger models can be designed in the future.

{
\ninept
\balance


\begin{thebibliography}{10}

\bibitem{hoppe1969beugung}
W~Hoppe,
\newblock ``Beugung im inhomogenen prim{\"a}rstrahlwellenfeld. iii.
  amplituden-und phasenbestimmung bei unperiodischen objekten,''
\newblock {\em Acta Crystallographica Section A: Crystal Physics, Diffraction,
  Theoretical and General Crystallography}, vol. 25, no. 4, pp. 508--514, 1969.

\bibitem{thibault2009probe}
Pierre Thibault, Martin Dierolf, Oliver Bunk, Andreas Menzel, and Franz
  Pfeiffer,
\newblock ``Probe retrieval in ptychographic coherent diffractive imaging,''
\newblock {\em Ultramicroscopy}, vol. 109, no. 4, pp. 338--343, 2009.

\bibitem{maiden2009improved}
Andrew~M Maiden and John~M Rodenburg,
\newblock ``An improved ptychographical phase retrieval algorithm for
  diffractive imaging,''
\newblock {\em Ultramicroscopy}, vol. 109, no. 10, pp. 1256--1262, 2009.

\bibitem{guizar2008phase}
Manuel Guizar-Sicairos and James~R Fienup,
\newblock ``Phase retrieval with transverse translation diversity: a nonlinear
  optimization approach,''
\newblock {\em Optics Express}, vol. 16, no. 10, pp. 7264--7278, 2008.

\bibitem{faulkner2004movable}
Helen Mary~Louise Faulkner and JM~Rodenburg,
\newblock ``Movable aperture lensless transmission microscopy: a novel phase
  retrieval algorithm,''
\newblock {\em Physical review letters}, vol. 93, no. 2, pp. 023903, 2004.

\bibitem{marchesini2007invited}
Stefano Marchesini,
\newblock ``Invited article: A unified evaluation of iterative projection
  algorithms for phase retrieval,''
\newblock {\em Review of scientific instruments}, vol. 78, no. 1, pp. 011301,
  2007.

\bibitem{elser2003phase}
Veit Elser,
\newblock ``Phase retrieval by iterated projections,''
\newblock {\em JOSA A}, vol. 20, no. 1, pp. 40--55, 2003.

\bibitem{fienup1982phase}
James~R Fienup,
\newblock ``Phase retrieval algorithms: a comparison,''
\newblock {\em Applied Optics}, vol. 21, no. 15, pp. 2758--2769, 1982.

\bibitem{gerchberg1972practical}
Ralph~W Gerchberg,
\newblock ``A practical algorithm for the determination of phase from image and
  diffraction plane pictures,''
\newblock {\em Optik}, vol. 35, pp. 237--246, 1972.

\bibitem{bunk2008influence}
Oliver Bunk, Martin Dierolf, S{\o}ren Kynde, Ian Johnson, Othmar Marti, and
  Franz Pfeiffer,
\newblock ``Influence of the overlap parameter on the convergence of the
  ptychographical iterative engine,''
\newblock {\em Ultramicroscopy}, vol. 108, no. 5, pp. 481--487, 2008.

\bibitem{zhong2016nonlinear}
Jingshan Zhong, Lei Tian, Paroma Varma, and Laura Waller,
\newblock ``Nonlinear optimization algorithm for partially coherent phase
  retrieval and source recovery,''
\newblock {\em IEEE Transactions on Computational Imaging}, vol. 2, no. 3, pp.
  310--322, 2016.

\bibitem{candes2015phase}
Emmanuel~J Candes, Xiaodong Li, and Mahdi Soltanolkotabi,
\newblock ``Phase retrieval via wirtinger flow: Theory and algorithms,''
\newblock {\em IEEE Transactions on Information Theory}, vol. 61, no. 4, pp.
  1985--2007, 2015.

\bibitem{ghosh2018adp}
Sushobhan Ghosh, Youssef~SG Nashed, Oliver Cossairt, and Aggelos Katsaggelos,
\newblock ``Adp: Automatic differentiation ptychography,''
\newblock in {\em 2018 IEEE International Conference on Computational
  Photography (ICCP)}. IEEE, 2018, pp. 1--10.

\bibitem{kandel2019using}
Saugat Kandel, S~Maddali, Marc Allain, Stephan~O Hruszkewycz, Chris Jacobsen,
  and Youssef~SG Nashed,
\newblock ``Using automatic differentiation as a general framework for
  ptychographic reconstruction,''
\newblock {\em Optics express}, vol. 27, no. 13, pp. 18653--18672, 2019.

\bibitem{nashed2017distributed}
Youssef~SG Nashed, Tom Peterka, Junjing Deng, and Chris Jacobsen,
\newblock ``Distributed automatic differentiation for ptychography,''
\newblock {\em Procedia Computer Science}, vol. 108, pp. 404--414, 2017.

\bibitem{rudin1992nonlinear}
Leonid~I Rudin, Stanley Osher, and Emad Fatemi,
\newblock ``Nonlinear total variation based noise removal algorithms,''
\newblock {\em Physica D: nonlinear phenomena}, vol. 60, no. 1-4, pp. 259--268,
  1992.

\bibitem{lefkimmiatis2013convex}
Stamatios Lefkimmiatis, Anastasios Roussos, Michael Unser, and Petros Maragos,
\newblock ``Convex generalizations of total variation based on the structure
  tensor with applications to inverse problems,''
\newblock in {\em International Conference on Scale Space and Variational
  Methods in Computer Vision}. Springer, 2013, pp. 48--60.

\bibitem{boominathan2018phase}
Lokesh Boominathan, Mayug Maniparambil, Honey Gupta, Rahul Baburajan, and
  Kaushik Mitra,
\newblock ``Phase retrieval for fourier ptychography under varying amount of
  measurements,''
\newblock {\em arXiv preprint arXiv:1805.03593}, 2018.

\bibitem{shamshad2019deep}
Fahad Shamshad, Farwa Abbas, and Ali Ahmed,
\newblock ``Deep ptych: Subsampled fourier ptychography using generative
  priors,''
\newblock in {\em ICASSP 2019-2019 IEEE International Conference on Acoustics,
  Speech and Signal Processing (ICASSP)}. IEEE, 2019, pp. 7720--7724.

\bibitem{heide2013high}
Felix Heide, Mushfiqur Rouf, Matthias~B Hullin, Bjorn Labitzke, Wolfgang
  Heidrich, and Andreas Kolb,
\newblock ``High-quality computational imaging through simple lenses,''
\newblock {\em ACM Transactions on Graphics (TOG)}, vol. 32, no. 5, pp. 1--14,
  2013.

\bibitem{kingma2014adam}
Diederik~P. Kingma and Jimmy Ba,
\newblock ``Adam: {A} method for stochastic optimization,''
\newblock in {\em 3rd International Conference on Learning Representations,
  {ICLR} 2015, San Diego, CA, USA, May 7-9, 2015, Conference Track
  Proceedings}, 2015.

\bibitem{paszke2019pytorch}
Adam Paszke, Sam Gross, Francisco Massa, Adam Lerer, James Bradbury, Gregory
  Chanan, et~al.,
\newblock ``Pytorch: An imperative style, high-performance deep learning
  library,''
\newblock in {\em Advances in Neural Information Processing Systems}, 2019, pp.
  8024--8035.

\bibitem{holler2019three}
Mirko Holler, Michal Odstrcil, Manuel Guizar-Sicairos, Maxime Lebugle,
  Elisabeth M{\"u}ller, Simone Finizio, Gemma Tinti, Christian David, Joshua
  Zusman, Walter Unglaub, et~al.,
\newblock ``Three-dimensional imaging of integrated circuits with macro-to
  nanoscale zoom,''
\newblock {\em Nature Electronics}, vol. 2, no. 10, pp. 464--470, 2019.

\bibitem{huang2014optimization}
Xiaojing Huang, Hanfei Yan, Ross Harder, Yeukuang Hwu, Ian~K Robinson, and
  Yong~S Chu,
\newblock ``Optimization of overlap uniformness for ptychography,''
\newblock {\em Optics Express}, vol. 22, no. 10, pp. 12634--12644, 2014.

\end{thebibliography}
}
\end{document}